\documentclass[preprint]{aastex}

\bibpunct{(}{)}{;}{a}{}{,} %% natbib format like A&A and ApJ
\usepackage{amsmath}

\shorttitle{Effect of a radiation cooling and heating on standing longitudinal oscillations}
\shortauthors{Kumar, Nakariakov \& Moon}

\begin{document}

\title{Effect of a radiation cooling and heating function on standing longitudinal oscillations in coronal loops}
\author{S. Kumar\altaffilmark{1}}
\email{sanjaykumar@khu.ac.kr}
\author{V. M. Nakariakov\altaffilmark{1,2,3}}
\and
\author{Y.-J. Moon\altaffilmark{1}}
\altaffiltext{1}{School of Space Research, Kyung Hee University, Yongin, 446-701, Gyeonggi, Korea}
\altaffiltext{2}{Centre for Fusion, Space and Astrophysics, Department of Physics, University of Warwick, Coventry CV4 7AL, UK}
\altaffiltext{3}{Central Astronomical Observatory of the Russian Academy of Sciences at Pulkovo, 196140 St Petersburg, Russia}

\begin{abstract}
Standing long-period (with the periods longer than several minutes) oscillations in large hot (with the temperature higher than 3 MK) coronal loops have been observed as the quasi-periodic modulation of the EUV and microwave intensity emission and the Doppler shift of coronal emission lines, {and have been interpreted as standing slow magnetoacoustic (longitudinal) oscillations. Quasi-periodic pulsations of shorter periods, detected in thermal and non-thermal emissions in solar flares could be produced by a similar mechanism.} We present theoretical modelling of {the standing slow magnetoacoustic mode,} showing that this mode of oscillation is highly sensitive to peculiarities of the radiative cooling and heating function. We generalised the theoretical model of standing slow magnetoacoustic oscillations in a hot plasma, including the effects of the radiative losses, and accounting for plasma heating. The heating mechanism is not specified and taken empirically to compensate the cooling by radiation and thermal-conduction. It is shown that the evolution of the oscillations is described by a generalised Burgers equation. Numerical solution of an initial value problem for the evolutionary equation demonstrates that different dependences of the radiative cooling and plasma heating on the temperature lead to different regimes of the oscillations, including growing, quasi-stationary and rapidly decaying. Our findings provide a theoretical foundation for probing the coronal heating function, and {may} explain the observations of decayless long-period quasi-periodic pulsations in flares. The hydrodynamic approach employed in this study should be considered with caution in the modelling of non-thermal emission associated with flares, as it misses potentially important non-hydrodynamic effects. 
\end{abstract}

\section{Introduction}
\label{sec:intro}

The main interest in magnetohydrodynamic (MHD) waves and oscillations in the solar atmosphere is connected with the possible role the waves play in heating of the atmospheric plasma, and with the exploitation of the plasma diagnostic potential of the waves \citep[see][for recent comprehensive reviews]{2012RSPTA.370.3193D,2014SoPh..289.3233L}. An important class of coronal MHD oscillations is standing waves in coronal loops, where structuring of the plasma across the magnetic field acts as a waveguide, and the loop footpoints are effective mirrors that form a standing wave pattern.

Standing longitudinal oscillations in coronal loops were discovered as a periodic Doppler shift of hot coronal emission lines with the SOHO/SUMER spectrometer \citep{2002ApJ...574L.101W}, and analysed in detail by \cite{2003A&A...406.1105W, 2003AN....324..340W}. The oscillations, usually called \lq\lq SUMER\rq\rq\ oscillations \citep[see][for a review]{2011SSRv..158..397W}, have periods in the range about 5--40 min, and are detected in long loops of the lengths of 200--300~Mm. SUMER oscillations are seen to be very rapidly decaying, with the decay time being about the period of oscillations. The decay time was found to be linearly proportional to the period. The relative amplitudes of plasma flows in SUMER oscillations reach 100--300~km~s$^{-1}$ that in some cases reaches 50\% of the sound speed corresponding to the temperature of the emitting plasma. Simultaneous observations of Doppler shift and intensity variations showed in some cases a quarter-period phase shift \citep{2005A&A...435..753W}. {About a half of the detected SUMER oscillations were found to be associated with flares \citep[e.g.][]{2011SSRv..158..397W}.}

Oscillations similar to the SUMER oscillations have been detected in the Doppler shift of hot coronal emission lines observed with Yohkoh/BCS \citep{2005ApJ...620L..67M, 2006ApJ...639..484M}.
Intensity oscillations with periods from 13 to 60 min in X-ray bright points were observed with Hinode/XRT by \citet{2011MNRAS.415.1419K}. Also, oscillations of soft X-ray intensity with periods of 12--30~min were found in the CORONAS-F/SPIRIT data \citep{2005ARep...49..579A}.
\cite{2012ApJ...756L..36K} observed a SUMER oscillation with the period 12.6 min and the decay time of 16 min simultaneously in the microwave and EUV emission, with the Nobeyama Radioheliograph and SDO/AIA, respectively. Similar oscillations, of the period about 32 min, were found in a white light curve of a stellar megaflare \citep{2013ApJ...773..156A}.
The first direct detection of SUMER oscillations in imaging data was recently reported by \cite{2013ApJ...779L...7K}. {Similar oscillatory patterns have recently been detected in the soft X-ray irradiance measurements data made with GOES \citep{2012ApJ...749L..16D, 2015SoPh..tmp...50S}.}

\cite{2002ApJ...580L..85O} were first to interpret SUMER oscillations in terms of standing slow magnetoacoustic oscillations of coronal loops. The oscillation period was found to be determined by the length of the loop and the sound speed. If finite-$\beta$ effects are accounted, the sound speed should be replaced by the tube speed. The dependence of the period of SUMER oscillations on the magnetic field in the finite-$\beta$ regime has been used for the seismological determination of the plasma parameter $\beta$ in the oscillating loop \citep{2007ApJ...656..598W}. Damping of the oscillations was associated with thermal conduction.

The simple numerical 1D model of \cite{2002ApJ...580L..85O} was further developed by \cite{2004A&A...414L..25N, 2004A&A...422..351T, 2004ApJ...605..493M, 2005A&A...438..713T, 2007ApJ...668L..83S, 2009A&A...495..313O} who subsequently included radiative effects, gravitational stratification, effects of the chromosphere near the footpoints, and effects of the magnetic field curvature. It was shown that standing slow magnetoacoustic oscillations are easily excited by a localised deposition of heat or increase in the plasma pressure. Also, the simulations showed that the oscillations may occur in two different regimes, the well-known rapidly decaying oscillations, and decayless oscillations. In the latter regime the oscillation is limited by the duration of the flare only. This regime has possibly been observed as the undamped oscillations of the flaring X-ray emission with a 20-min period detected by \cite{1994SoPh..152..505S}; a 143-s period detected by \cite{2002AstL...28..397T};
{the 25 to 48~s pulsations detected in the hard X-rays in the initial phase of three flares by
\cite{2003SoPh..218..183F}}; the 60-s variations of the H$\alpha$ emission detected by \cite{2005SoPh..229..227H}; the long-period ($\ge 60$~s) variations of the radio and X-ray fluxes, detected by \cite{2006A&A...460..865M}; the 5-min and 13.5-min oscillatory modulation of the 8-mm emission, revealed by \cite{2006SoPh..233...89K}; the persistent, semi-regular compressions of the flaring core region, modulating the plasma temperature and emission measure with the period of about 60~s, detected in soft X-rays and EUV by \cite{2013ApJ...777..152S}; {and the 4-min pulsations of the hard X-ray, radio and EUV emissions in a flare, detected by \cite{2015ApJ...807...72L}. }

{For the theoretical analysis of waves, a useful alternative to full-scale numerical simulations is the method of an evolutionary equation. In this method, the wave evolution, e.g. its damping or amplification, wave shape deformation, and acceleration, is determined in terms of the intrinsic evolutionary mechanisms, such as dissipation, dispersion, nonlinearity and activity of the medium. These evolutionary mechanisms are modelled by specific terms in the evolutionary equation. Specific expressions for the coefficients in front of these terms can be determined from a certain set of governing equations, e.g. MHD, or can be taken in a guessed, effective form, e.g. when some necessary but not understood physical processes (such as coronal heating) should be included in the model. Moreover, these coefficients could be determined empirically, from observations and then used to constrain the guessed expressions.
In the coronal context, this approach has turned to be successful in modelling wave phenomena in the corona, e.g. propagating longitudinal waves in coronal active region fans \citep{2004A&A...414L..25N, 2004A&A...422..351T, 2015A&A...573A..32A} and polar plumes \citep{2000ApJ...533.1071O}, nonlinear Alfv\'en waves in coronal holes \citep{2000A&A...353..741N}.}

{For standing longitudinal waves in coronal loops, the evolutionary equation method was recently used by \citet{2013A&A...553A..23R} (referred to as R13 in the following discussion). The model designed in R13} is based on the asymptotic expansion with the use of the small parameter, determined by the weakness of the effects of nonlinearity and dissipation by finite thermal conduction and/or viscosity. It was shown that in this approximation the SUMER oscillation can be considered as a superposition of two oppositely-propagating nonlinear waves governed by the Burgers equation. An interesting consequence of this study was a periodic movement of the position of the highest amplitude along the loop, caused by the nonlinearity.

An important effect that needs to be accounted for in the description of standing longitudinal oscillations in coronal loops is the radiation from the perturbed plasma. The radiation is mainly controlled by the composition of the plasma and the presence of heavy, not fully ionised ions. In the context of longitudinal oscillations the effect of radiative losses was analytically described by \cite{2013Ge&Ae..53.1013B}, and shown to lead to enhanced damping.
Dependence of the radiative losses on the plasma temperature and pressure is quite non-monotonic, and includes segments with both positive and negative gradients, see, e.g. Fig.~1 of \cite{2009A&A...508..751S}, and is additionally modified in the presence of heating. It has been known for long time \citep[e.g.][]{1965ApJ...142..531F} that under certain circumstances, e.g. in the presence of heating,
thermal instability can occur in a diffuse medium due to imbalance between temperature-independent energy gains, i.e., heating, and temperature-dependent radiative losses. From the point of view of magnetoacoustic wave dynamics, peculiarities of the energy loss/gain function dependence on thermodynamical parameters (e.g., density and pressure) may lead to the amplification of oscillations and hence an increase in the nonlinearity \citep[e.g.][]{2000ApJ...528..767N}. The balance of the radiative/heating effects and dissipation may lead to appearance of stationary propagating nonlinear waves (autowaves) of a saw-tooth shape \citep{2010PhPl...17c2107C,2011Ap&SS.334...35M}, slow magnetoacoustic auto-solitons \citep{1999PhLA..254..314N}, and nonlinear resonant amplification of Alfv\'en waves \citep{2014TePhL..40..701Z}.

The aim of this paper is to generalise the work of \cite{2013A&A...553A..23R} accounting for isentropic effects based on the presence of an energy loss/gain function in the energy equation. In Sec.~\ref{goveq} we discuss the model and governing equations. In Sec.~\ref{disrel} we derive and analyse linear dispersion relations. In Sec.~\ref{seceveq} we derive the nonlinear evolutionary equation. In Sec.~\ref{evolution} we study different regimes of the oscillations. The results obtained are summarised in Sec.~\ref{disc}.

\section{Governing equations}
\label{goveq}

In this study we ignore 2D effects, such as the loop curvature and transverse non-uniformity, and consider longitudinal oscillations as field-aligned acoustic oscillations. Effects of stratification are neglected too, and the loop is taken to be situated between 0 and $L$ along the magnetic field directed along the $z$ axis.
The governing set of equations is the continuity, Euler and energy equations,
\begin{eqnarray}
\frac{\partial \rho}{\partial t} + \frac {\partial (\rho u)}{\partial z} &=& 0,\label{goveq01} \\
\frac{\partial u}{\partial t} + u{ \frac {\partial  u}{\partial z}} &=& -\frac{1}{\rho}\frac{\partial p}{\partial z}
+\frac{1}{\rho} {\frac{\partial}{\partial z}}\rho \nu {\frac{\partial u}{\partial z}}, \label{goveq02} \\
\frac{\partial p}{\partial t} -\frac{\gamma p}{\rho}{ \frac {\partial   \rho}{\partial t}} &=& (\gamma - 1)[Q(\rho, p)+ \nabla( \kappa \nabla T)], \label{goveq03} \\
p &=& \frac{k_\mathrm{B}}{m}T\rho,
\label{goveq04}
\end{eqnarray}
where $\rho$ is the plasma density; $T$ is the temperature; $p$ is the plasma pressure; $u$ is the speed of a field-aligned bulk flow; $\gamma$ is the adiabatic index; $t$ is the time; the coefficients
\begin{eqnarray}
 \nu = \frac{4 \eta_0}{3 \rho}, \ \mbox{and}\ \  \kappa = \frac {(\gamma-1) m \kappa_\parallel}{\rho k_\mathrm{B}}
\end{eqnarray}
describe the viscosity and field-aligned thermal conduction, respectively, determined by the coefficients of the bulk viscosity $\eta_0$ and thermal conductivity $\kappa_\parallel$; and  $Q(\rho, p)$ is the cooling/heating (the energy loss/gain) function that accounts for radiative cooling and unspecified coronal heating.
In this study the coefficients of the bulk viscosity $\eta_0$ and thermal conductivity $\kappa_\parallel$ are not specified as they are likely to be enhanced by microturbulent processes typical for plasmas. Eq.~(\ref{goveq03}) extends the governing equations used in R13 by including the cooling/heating function $Q(\rho, p)$.

In this study we consider weak perturbations of the equilibrium, determined by the constant equilibrium density $\rho_0$, pressure $p_0$, temperature $T_0$ that are assumed to be uniform along the loop. Hence in the equilibrium thermal conduction is zero. Thus, in the equilibrium the cooling/heating function $Q(\rho_0, p_0)=0$, i.e. the heating compensates the radiative losses. The same equilibrium was considered, e.g., in \cite{2000ApJ...528..767N, 2010PhPl...17c2107C,2011Ap&SS.334...35M}.
The perturbations of the equilibrium have a form of field-aligned flows that satisfy the
boundary conditions
\begin{equation}
                u=0\   \mbox{at}  \      z=0,    L,
\label{bc}
\end{equation}
i.e. the chromosphere is considered to be a rigid wall for the longitudinal oscillations.

The perturbation amplitude is characterised by a dimensionless small parameter $\epsilon \ll 1$. The dissipative effects caused by finite thermal conduction and viscosity, and non-adiabatic effects caused by the radiative losses and heating, are considered to be of the order of $\epsilon$ too. Thus, the evolutionary equation will include quadratically nonlinear terms together with linear terms that represent the non-adiabatic processes.
It is convenient to introduce the scaled coefficients at the dissipative terms as follows
\begin{equation}
 \overline\nu = \epsilon^{-1}  \nu \ \  \mbox{and} \ \ \overline{\kappa} = \epsilon^{-1} \kappa.
\label{eq5}
\end{equation}

\section{Dispersion relations}
\label{disrel}

Linearising the set of equations (\ref{goveq01}--\ref{goveq04}), we obtain
\begin{eqnarray}
\frac{\partial \rho}{\partial t} + \rho_0 \frac {\partial  u}{\partial z} &=& 0, \label{lin1} \\
\frac{\partial u}{\partial t} + \frac{1}{\rho_0}\frac{\partial p}{\partial z} &=&
\overline{\nu} {\frac{\partial^2 u}{\partial z^2}}, \label{lin2} \\
\frac{\partial p}{\partial t} -\frac{\gamma p_0}{\rho_0} { \frac {\partial  \rho}{\partial t}} &=& (\gamma - 1)\Big[\tilde{a_p} p + \tilde {a_\rho} \rho + \overline{\kappa} \frac {\partial^2 T}{\partial z^2}\Big], \label{lin3} \\
p - \frac{k_B \rho_0}{m}T - \frac{k_B  T_0}{m} \rho&=&0,
\label{lin4}
\end{eqnarray}
where we used the linear terms in the Taylor expansion of the cooling/heating function $Q$ near the equilibrium, with $\tilde{a_\rho}={\partial Q}/{\partial \rho}$ taken at $p_0$, and $\tilde{a_p}={\partial Q}/{\partial p}$ taken at $\rho_0$. Here, the variables $p$, $\rho$, $u$ and $T$ are perturbations of the equilibrium state. In the following consideration we assume that the terms on the righthand side are assumed to be smaller than the terms on the lefthand side.

Assuming the harmonic dependence of the perturbed quantities $\sim \exp(-i\omega t + i k z)$ we obtain the dispersion relation
\begin{equation}
\omega^2-C_\mathrm{s}^2 k^2-  \frac{i(\gamma - 1)}{\rho_0} \Big(\frac{\rho_0  \tilde{a_\rho} k^2}{\omega} + \tilde{a_p} \rho_0 \omega
- \frac{\overline{\kappa}  m  \omega  k^2}{k_B}+\frac{\overline{\kappa} T_0 k^4}{\omega}\Big) +i\overline{\nu}\omega k^2 = 0 ,
\label{disprel}
\end{equation}
where $\omega$ is the cyclic frequency, $k$ is the wave number, and $C_\mathrm{s} = (\gamma p_0/\rho_0)^{1/2}$ is the sound speed.

Taking that the lefthand side terms in Eqs.~(\ref{lin2}) and (\ref{lin3}) are small, and hence the last two terms in  Eq.~(\ref{disprel}) are smaller than the first two terms, we obtain that
 $ \omega \approx C_\mathrm{s} k$, and determine the real and imaginary parts of the frequency as
\begin{eqnarray}
 \omega_\mathrm{R}&\approx&C_\mathrm{s} k, \label{omegar} \\
 \omega_\mathrm{I} &\approx& \frac{(\gamma - 1)}{2}A
 -\Big[\frac{(\gamma - 1)^2 m\overline{\kappa}}{2\gamma \rho_0 k_B}+ \frac{\overline{\nu}}{2}\Big]k^2, \label{omegai}
\end{eqnarray}
respectively, where
\begin{equation}
A= \tilde{a_\rho}/C_\mathrm{s}^2+\tilde{a_p}
\label{adef}
\end{equation}
is the parameter determined by the heating/cooling function $Q$.

Eq.~(\ref{omegar}) shows that the oscillation period is determined by the length of the loop, e.g. $P = 2\pi/ \omega_\mathrm{R} = \pi/C_\mathrm{s}L$ for the fundamental longitudinal mode, and the sound speed. Eq.~(\ref{omegai}) contains two terms. The second term on the righthand side, which contains the thermal conductivity $\overline{\kappa}$ and viscosity $\overline{\nu}$ is always negative, and hence causes damping of the oscillation. The damping time is inversely proportional to $k^2$, thus  oscillations in shorter loops decay more rapidly. The first term on the righthand side of (\ref{omegai}) can be either positive or negative, depending at the local gradients of the cooling/heating function, $A$.

In the case $A < 0$ this terms contributes to damping. However, as it is independent of $k$ the damping caused by this term is the same in short and long loops. In the case $A > 0$, this term suppresses damping. When the condition
\begin{equation}
A   = A_\mathrm{crit} = \frac{2}{(\gamma - 1)} \Big[\frac{(\gamma - 1)^2 \overline{\kappa} m}{2\gamma \rho_0 k_B}+ \frac{\overline{\nu}}{2}\Big]k^2
\label{acrit}
\end{equation}
is fulfilled, the oscillation becomes undamped. For $A > A_\mathrm{crit}$, the plasma becomes unstable and the oscillation amplitude grows in time --- the thermal over-stability occurs. The critical value $A_\mathrm{crit}$ corresponds to the threshold of the over-stability, which is determined by thermodynamical parameters of the plasma, the heating/cooling function, and the length of the loop.

\section{Evolutionary equation for standing longitudinal oscillations}
\label{seceveq}

Observations show that the relative amplitude of standing longitudinal oscillations in coronal loops reaches 30--50\% {\citep[e.g.][]{2011SSRv..158..397W}}. Thus, it is necessary to account for nonlinear effects in the evolution of the oscillations.
The presence of the small parameter allows us to perform the asymptotic analysis of weakly-nonlinear, weakly-isentropic standing longitudinal oscillations, following the methodology developed in R13.

Consider the nonlinear and non-adiabatic processes (the latter are caused by the finite viscosity and thermal conductivity, and the heating/cooling function) to operate at the slow time
 $t_1=\epsilon t$. Thus, we look for a solution to Eqs.~(\ref{goveq01}--\ref{goveq04}) in the form of expansions
\begin{equation}
f=f_0 +\epsilon f_1+\epsilon^2 f_2+...
\label{expan}
\end{equation}
where $f$ represents the quantities $u$, $\rho$, $p$ and $T$.  The term $f_0$ represents the unperturbed state  i.e.  $f_0 = \mathrm{const}$ with $u_0 =0$.

\subsection{First-order approximation}

Substituting the expansions (\ref{expan}) into Eqs.~(\ref{goveq01})--(\ref{goveq04}) we collect the terms of the same power of the small parameter $\epsilon$. The first order approximation, after the elimination of all variables in favour of $u_1$ gives us the acoustic wave equation,
\begin{equation}
 \frac{\partial^2 u_1}{\partial t^2} - C_\mathrm{s}^2  \frac{\partial^2 u_1}{\partial z^2}  = 0
\label{order1}
\end{equation}
Eq.~(\ref{order1}) has the solution $u_1 = C_\mathrm{s}\left[f(\xi) + g(\eta)\right]$,  where
\begin{equation}
\xi = \omega(t-{z}/{C_\mathrm{s}}),\ \eta = \omega(t+{z}/{C_\mathrm{s}})
\label{vars}
\end{equation}
are dimensionless variables; and $f(\xi)$ and $g(\eta)$ are arbitrary smooth functions that describe the waves travelling in the positive and negative directions $z$, respectively.

Applying the boundary conditions given by Eq.~(\ref{bc}) we obtain the solution in a form of standing waves,
\begin{equation}
u_1 = C_\mathrm{s} [f(\xi) - f(\eta)],
\label{stand1}
\end{equation}
which is a superposition of two waves propagating in opposite directions. The function $f(y)$ is periodic with the period $2\pi$, which requires $\omega = \pi C_\mathrm{s}/L$. This cyclic frequency corresponds to the fundamental longitudinal mode of a loop of length $L$, filled in with a uniform plasma with the sound speed $C_\mathrm{s}$. The perturbations of other physical quantities are
\begin{equation}
\rho_1=\rho_0[f(\xi) + f(\eta)],\      p_1=\rho_0 C_\mathrm{s}^2[f(\xi) + f(\eta)],\
 T_1=(\gamma - 1)T_0[f(\xi) + f(\eta)].
 \label{stand2}
 \end{equation}
Solutions (\ref{stand1})--(\ref{stand2}) correspond to the real part of the solution to dispersion relation (\ref{omegar}).

\subsection{Second-order approximation}

Collecting terms of the order of $\epsilon^2$, and again eliminating all variables in favour of $u_2$, we obtain
\begin{eqnarray}
\label{order2}
\frac{\partial^2 u_2}{\partial t^2}-C_\mathrm{s}^2  \frac{\partial^2 u_2}{\partial z^2} &=&  \frac{\omega^3[\gamma \overline{\nu}+(\gamma - 1)\overline{\kappa}]}{\gamma C_\mathrm{s}}(f'''_--f'''_+)
-2 \omega  C_\mathrm{s}(\frac{\partial f'_-}{\partial t_1}-\frac{\partial f'_+}{\partial t_1}) \nonumber \\
&+&\omega^2 C_\mathrm{s}\Big[ (\gamma + 1)\big({f'_-}^2 - {f'_+}^2 + f_{-}f''_-  - f_{+}f''_+) \nonumber \\
&+&(3-\gamma)(f_{-}f''_+  - f_{+}f''_-)\Big] +(\gamma-1){\omega}{C_\mathrm{s}}A(f'_- -f'_+) , \nonumber \\
\end{eqnarray}
where
$f_{-}= f(\xi, t_1)$, $f_{+} = f(\eta, t_1)$, and the prime denotes the partial derivatives of the function $f(y, t_1)$ with respect to the spatial coordinate $y$, the independent variable $t_1$ is the \lq\lq slow\rq\rq\ time that describes the evolution of the oscillations in the presence of the effects of non-adiabaticity and nonlinearity (i.e. the righthand side of Eq.~(\ref{order2})). Eq.~(\ref{order2}) is similar to Eq.~(30) of R13 with the additional term on the righthand side, which accounts for the cooling/heating function via the parameter $A$. Eq.~(\ref{order2}) should be supplemented by the boundary conditions
\begin{equation}
                u_2=0\   \mbox{at}  \      z=0,    L.
\label{bc2}
\end{equation}

The asymptotic expansion given by Eq.~(\ref{expan}) is valid when the second order terms $f_2$ do not experience a secular growth. Such a growth is possible if the righthand side terms of Eq.~(\ref{order2}) are in resonance with the eigenfunctions of the boundary problem constituted by the lefthand side of Eq.~(\ref{order2}) and boundary conditions (\ref{bc2}). This possibility is excluded by demanding the righthand side of Eq.~(\ref{order2}) to be orthogonal to the eigenfunctions of Eq.~(\ref{order2}) with (\ref{bc2}). Following the procedure described in R13 \citep[similar methods have also been used in the solar context in the description of weakly-nonlinear fast waves in][]{1995SoPh..160..289N,1997JPlPh..58..315N} we obtain the condition of the orthogonality,
\begin{equation}
\frac{\partial f}{\partial \tau}-2\lambda f \frac{\partial f}{\partial y}-\frac{\partial^2 f}{\partial y^2}-\alpha f=0,
\label{eveq}
\end{equation}
where $\tau={t}/{t_\mathrm{dl}}$, and
\begin{equation}
t_\mathrm{dl}= \frac{2\gamma L^2}{\pi^2[\gamma\nu+(\gamma-1)\kappa]},\
\lambda=\frac{\epsilon \gamma(\gamma+1)C_\mathrm{s} L}{2 \pi [\gamma\nu+(\gamma-1)\kappa]},\ \alpha= \frac{\epsilon \gamma(\gamma-1)A L^2}{\pi^2 [\gamma\nu+(\gamma-1)\kappa]}.
\label{eveqpar}
\end{equation}
Eq.~(\ref{eveq}) is a generalised Burgers equation, similar to Eq.~(38) of R13, with the fourth, linear term accounting for the finite cooling/heating function. Solutions to Eq.~(\ref{eveq}) with boundary conditions (\ref{bc}) should be substituted in Eq.~(\ref{stand1}) that describes a standing oscillation and its evolution.

\section{Evolution of standing longitudinal waves}
\label{evolution}

\begin{figure}[htp]
  \centering
  \begin{tabular}{cc}
    \includegraphics[width=60mm]{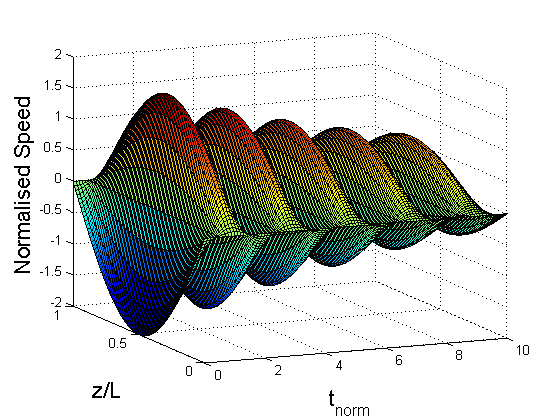}&
    \includegraphics[width=60mm]{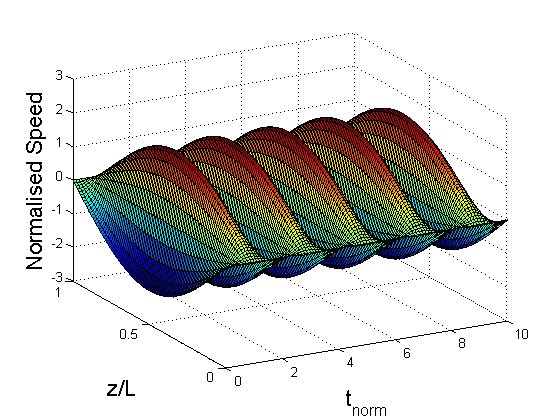}\\
    \includegraphics[width=60mm]{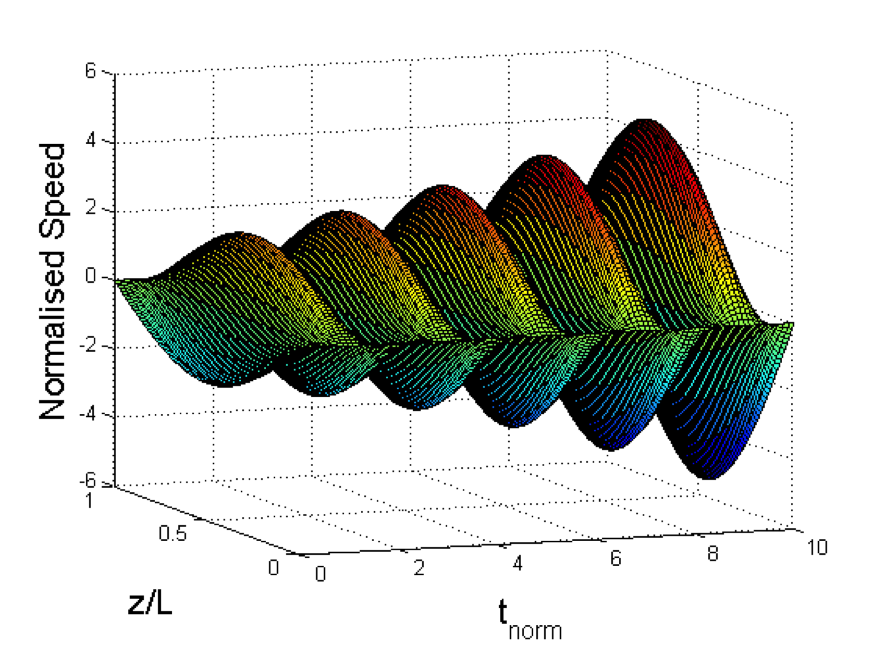}&
    \includegraphics[width=60mm]{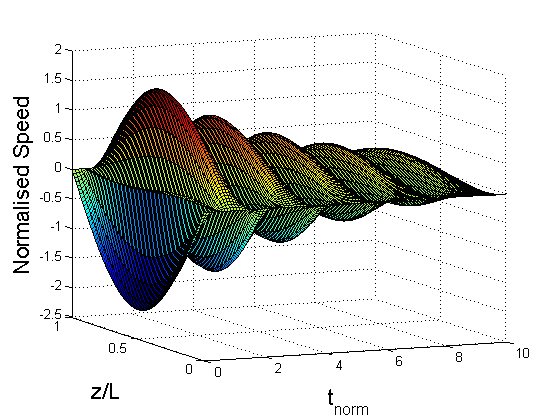}\\
    \includegraphics[width=60mm]{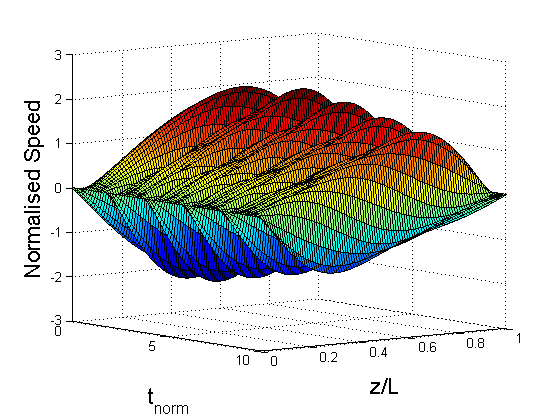}&
    \includegraphics[width=60mm]{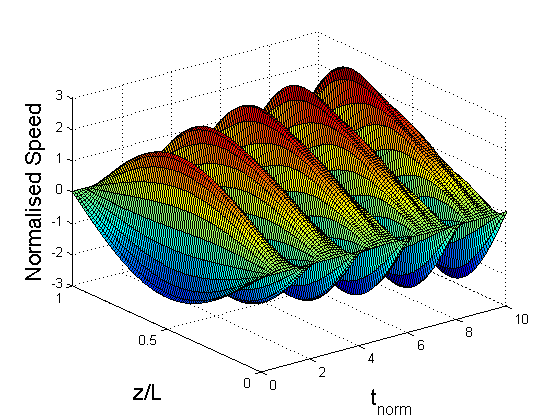}
  \end{tabular}
 \caption{Different regimes of the evolution of the global longitudinal oscillation of a coronal loop, determined by different combinations of the parameters of nonlinearity and non-adiabaticity. The plasma speed is normalised at double the initial amplitude. The time is normalised at half the linear oscillation period. The spatial coordinate is normalised at the loop length $L$.
The top raw: the left panel shows a decaying linear oscillation ($\lambda = 0$, $\alpha=0$); the right panel shows an almost undamped linear oscillation ($\lambda = 0$, $\alpha=10$).
The middle raw: left panel shows a growing oscillation ($\lambda = 0$, $\alpha=20$);
the right panel shows an over-damped linear oscillation ($\lambda = 3$, $\alpha=-10$).
Bottom raw: the left panel shows the an undamped highly-nonlinear oscillation ($\lambda = 6$, $\alpha=10$); the right panel shows a growing nonlinear oscillation ($\lambda = 3$, $\alpha=15$). }

    \label{f1}
\end{figure}

\begin{figure}[htp]
  \centering
  \begin{tabular}{cc}
    \includegraphics[width=60mm]{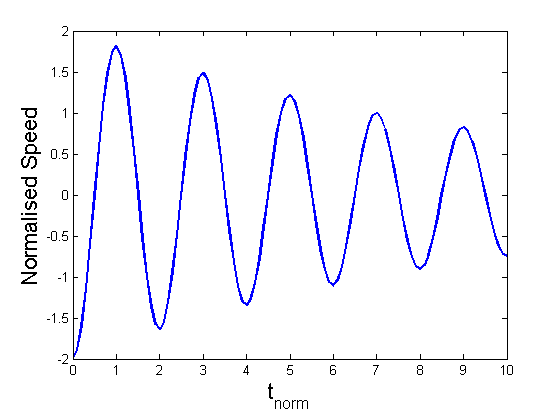}&
    \includegraphics[width=60mm]{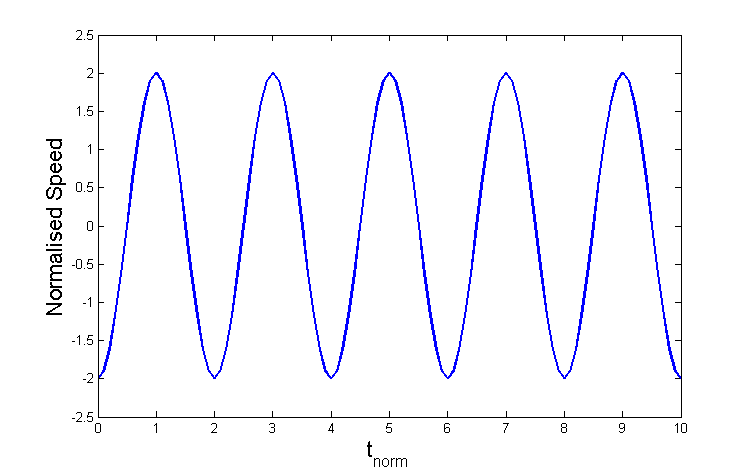}\\
    \includegraphics[width=60mm]{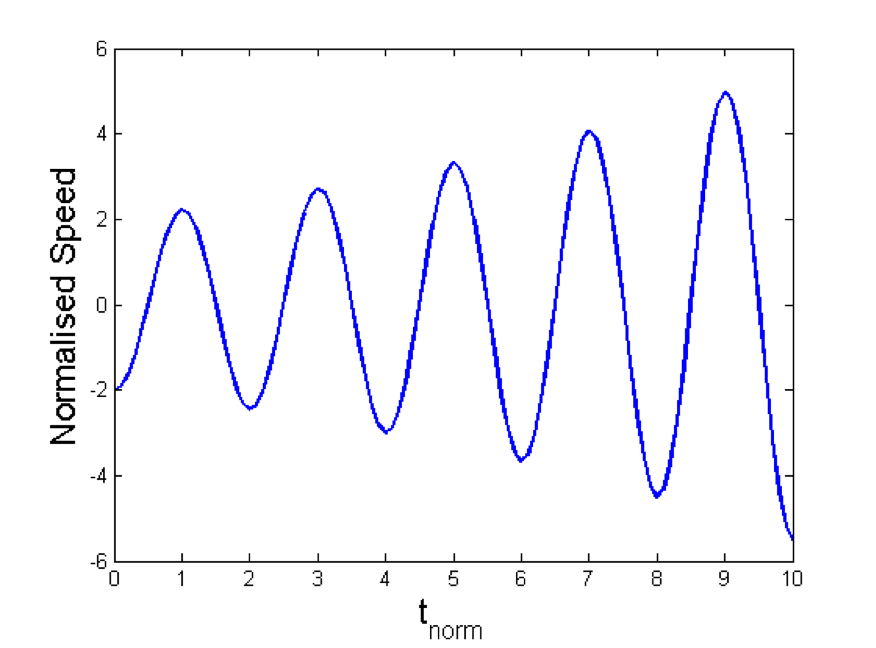}&
    \includegraphics[width=60mm]{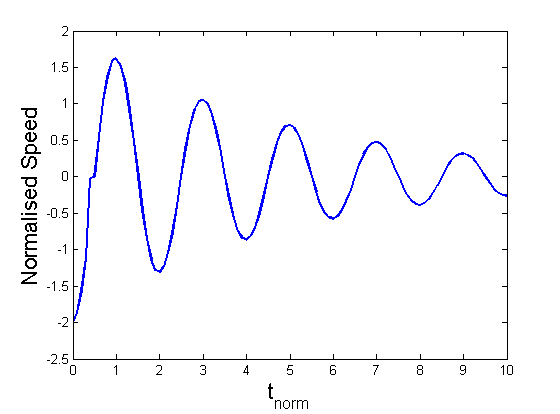}\\
    \includegraphics[width=60mm]{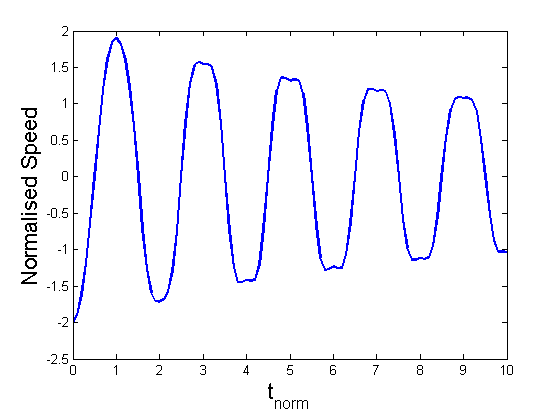}&
    \includegraphics[width=60mm]{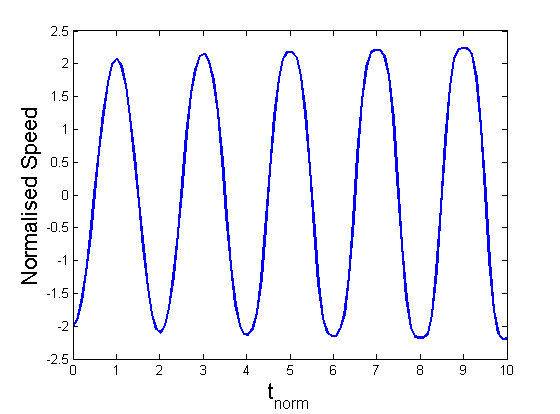}
  \end{tabular}
  \caption{Variations of the longitudinal speed at the top of the loop in different regimes of the evolution of the global longitudinal oscillation, determined by different combinations of the parameters of nonlinearity and non-adiabaticity. The plasma speed is normalised at double the initial amplitude. The time is normalised at half the linear oscillation period. The panels correspond to the regimes shown in the corresponding panels of Figure~\ref{f1}.}
  \label{f2}
\end{figure}

An initial value problem constituted by Eqs.~(\ref{eveq}),  (\ref{bc}), (\ref{stand1}), and the initial condition $f(y, 0) = - \sin(y)$ was solved numerically with the use of the standard procedure \textit{pdede} in Matlab~8.5\footnote{mathworks.com/products/matlab/}. Numerical solutions $f(y,t_1)$ with the use of expressions (\ref{vars}) were substituted in equation (\ref{stand1}) to obtain the oscillations of the field-aligned velocity $u_1$.

Figures~\ref{f1} and \ref{f2} show different regimes of the oscillations, determined by different combinations of the parameters of Eq.~(\ref{eveq}). The top-left panels of both the figures show the decaying linear oscillation that was in detail considered in R13. Other panels show the effect of the cooling/heating function $Q(\rho,p)$. In the middle-left panels the oscillation grows because of the thermal over-stability.
In the top-right the oscillations is undamped. This regime occurs when the profile of the dependence of the cooling/heating function on the density and pressure has a value given by condition (\ref{acrit}). In this case the damping caused by thermal conduction and viscosity is compensated by the amplification caused by the thermal over-stability. In the middle-right panels the oscillation damps stronger than in the case without the cooling/heating function (the top-left panel). In this case the profile of $Q(\rho,p)$ does not satisfy condition (\ref{acrit}), and hence contribute to the oscillation damping. Despite the finite value of the nonlinear coefficient $\lambda$, the oscillation remains practically harmonic, as it rapidly decreases in time. The bottom panels show nonlinear deformation of the oscillations caused by the finite amplitude. The plot in the bottom-left panel is rotated to make the nonlinear deformation better visible. The nonlinearity manifests itself as the appearance of higher spatial harmonics of the oscillation, i.e. the instant snapshots of the oscillations can be considered as a sum of the fundamental mode $\sin(\pi z/L) $ and the nonlinearly generated modes $\sin(\pi N z/L)$, where the integer $N\ge 2$ is the spatial harmonic number. Fig.~\ref{f2} shows the gradual distortion of the oscillation profile that becomes anharmonic. For higher values of the nonlinear coefficient $\lambda$ (c.f. the bottom-left and bottom-right panels), the distortion is more pronounced. Observationally, it would lead to the movement of the position of the highest amplitude along the loop from the top where it is in the fundamental mode.

\section{Discussion and conclusions}
\label{disc}

Our study shows that the cooling/heating function that accounts for radiation and unspecified heating of the coronal plasma can significantly affect longitudinal (slow magnetoacoustic) oscillations of coronal loops.
The main contribution to this effect is caused by the gradient of this function at the equilibrium point on the thermodynamic parametric plane (e.g. the $p$--$\rho$ plane). The specific dependence of the cooling/heating function on these parameters has not been established yet, as it is connected with the enigmatic coronal heating mechanism.
Moreover, fine details of the radiative-loss function are also continuously updated following new and improved calculations of atomic data and transition rates \citep[see, e.g., discussion in][]{2012A&A...543A..90R}. What is also important for our study is that the dependence of the radiative-loss function on the thermodynamical parameters is not steady, and even approximated dependences show steep positive and negative gradients for coronal conditions. Thus, we could treat the cooling/heating function as a free parameter in our study. Unfortunately, this uncertainty does not allow us to make any quantitative estimations, restricting our attention to the discussion of the possible regimes and their seismological implications only.

It should be pointed out that governing equations (\ref{goveq04}) used in our derivation, as well as in R13, are rather simple, and may miss some important physical effects. In particular, these additional effects include the complex interactions between thermal and non-thermal effects in flares, and long-durational, in comparison with the oscillation period, field-aligned up- and downflows \citep[e.g.][]{2003A&A...399.1159F, 2004ApJ...611L..49W, 2015ApJ...811....7L}.
Thus, the specific values of the coefficients given by Eqs.~(\ref{eveqpar}) may need to be modified if these additional effects are taken into account. However, despite the possible changes in the governing equations, the general view of the evolutionary equation will be similar to Eq.~(\ref{eveq}) that accounts for the intrinsic mechanisms responsible for the wave evolution: nonlinearity, dissipation and activity. Our main finding is that the effects associated with the activity of the medium, modelled by the fourth term in Eq.~(\ref{eveq}), may cause a dramatic change in the slow wave evolution, and should not be neglected. Furthermore, in the context of the the specific value of the radiative losses is not important, as the effect of the thermal over-stability is prescribed by the derivatives of the radiative losses and heating function with respect to the local thermodynamical parameters of the plasma.

Anyway, it is clear that one of obvious shortcomings of the presented analysis is the applicability of the effective fluid approach to flaring plasmas, which is not established and needs a dedicated study.
On the other hand, the Burgers equation formalism is known to work well as the zero-order approximation even in a collisionless plasma \citep[e.g.][]{1975SVPCS...8.....H}. Also, the main intrinsic features of nonlinear wave dynamics, such as nonlinear cascade and appearance of dissipative structures, described by the generalised Burgers equation given by expression~(\ref{eveq}) are similar in very different environments.

In addition, the formalism developed in this study may be applied to the loops surrounding the flaring site, where the applicability of the fluid approach is justified. In this case, the oscillatory modulation of thermal emission (e.g. EUV, soft X-ray) come from the variation of the plasma density and temperature.
Oscillatory modulation of non-thermal emission (e.g. microwave, hard X-ray, $\gamma$-ray) could be produced by the modulation of the magnetic reconnection rate by a magnetoacoustic oscillation in a loop situated nearby the flaring site \citep{2006SoPh..238..313C, 2006A&A...452..343N}. In those scenarios, the modulation of non-thermal emission is produced by the periodic modulation of the plasma resistivity, caused by the modulation of the macroscopic plasma parameters, such as the density and temperature, by an MHD or acoustic wave. In addition, the periodic modulation of the density of the plasma, and hence the electron plasma frequency, by a slow magnetoacoustic wave, can periodically modulate the gyrosynchrotron emission produced by non-thermal electrons \citep{2006A&A...446.1151N}. Another possibility for the modulation of the non-thermal emission by a periodic variation of macroscopic plasma parameters in an MHD wave is the periodic variation of the magnetic mirror condition in the legs of flaring loops \citep{1982SvAL....8..132Z}. Thus, admitting that the non-thermal emission is definitely caused by non-MHD effects, we would like to point out that its periodic modulation can be associated with the periodic variations of the macroscopic plasma parameters in MHD oscillations, considered in this paper.

We found that, depending on the specific gradient of the cooling/heating function at the thermal equilibrium there are three main different regimes of longitudinal oscillations possible in coronal loops. The radiative cooling and heating effects can either increase the oscillation damping, or suppress the damping caused by finite thermal conduction and viscosity. In the latter case we can observe either undamped oscillations, or even increase in the oscillation amplitude in time {- the regime of thermal over-stability}. In all these regimes, the oscillation period remains determined by the loop length and temperature. We should point out that undamped oscillations have been detected in coronal oscillations during solar flares \citep[e.g.][]{1994SoPh..152..505S, 2002AstL...28..397T, 2003SoPh..218..183F, 2005SoPh..229..227H, 2006A&A...460..865M, 2006SoPh..233...89K, 2013ApJ...777..152S, 2015ApJ...807...72L}, and hence could, at least in some cases, be attributed to this effect. In the undamped regime, the oscillation period remains determined by the length of the loop and the temperature of the plasma. For example, for a 120-s oscillation in a flaring plasma of 20~MK temperature, the length of the oscillating loop should be about 40~Mm for the fundamental harmonics, and 80~Mm for the second spatial harmonics. It is necessary to mention that an undamped or growing regime of another MHD mode, the kink oscillation, has recently been discovered observationally \citep{2012ApJ...751L..27W, 2013A&A...552A..57N} during non-flaring periods of time, while its nature remains unrevealed. Anyway, undamped kink oscillations are not likely to be responsible for the undamped or growing QPP detected in solar flares, as that regime is observed during the quiet periods of the solar activity.

In contrast with the damping caused by finite thermal conduction and/or viscosity that decreases with the oscillation wavelength, the cooling/heating function is independent of the wavelength. It suggests that the undamped and over-stable regimes are more likely to occur in longer loops, {in which the efficiency of the damping by thermal conduction and viscosity is lower}. However, realisation of these regimes in specific situations depends on the specific thermodynamical parameters of the plasma in the oscillating loop. Also, the observational detection of an undamped or growing long-period oscillation is only possible in the case when the flaring emission in e.g. the soft X-rays lasts longer than several {cycles of the oscillation}.

The nonlinear movement of the position of the highest amplitude along the loop, that was revealed in R13, becomes even more important in the case of undamped or growing oscillations. In those cases, nonlinear corrections get accumulated for a longer time, causing more significant departure from the harmonic shape of the oscillations.

Thermal over-stability can also lead to the excitation of oscillations. A gradual change of thermodynamical conditions in a loop, could reach the instability's threshold (\ref{acrit}),
causing the onset of the over-stability and hence increase in the oscillation amplitude. In this reasoning one could also take into account the possible onset of some plasma micro-instabilities caused, e.g. by plasma flows in the oscillations, resulting in the increase in the viscosity and thermal conductivity. This scenario could explain the sudden appearance of the oscillation and its rapid decay by the enhance dissipation. However, this discussion remains speculative till more detail investigation of this possibility.

Our results demonstrate that the behaviour of slow magnetoacoustic oscillations in coronal loops is sensitive to the peculiarities of the coronal cooling/heating function. Different dependences of the combination of the radiative cooling and heating on the plasma's thermodynamical parameters result in qualitatively different regimes of the oscillations (over-damped, undamped, growing), providing us with a potential ground for the seismological diagnostics of the cooling/heating function in observations. This finding motivates a more detailed study of compressive oscillations in observational data. Special attention should be paid to the search for the undamped and growing regimes, similar to those described in \citep[e.g.][]{1994SoPh..152..505S, 2002AstL...28..397T,  2003SoPh..218..183F, 2005SoPh..229..227H, 2006A&A...460..865M, 2006SoPh..233...89K, 2013ApJ...777..152S, 2015ApJ...807...72L}, the shape of the oscillation curve, and the appearance of higher spatial harmonics.

{The significant limitations of the used governing equations, discussed above, require a further development of the model by including additional physical effects typical for flaring plasmas. The formalism for the derivation of evolutionary equation (\ref{eveq}) presented here provides one with a convenient starting point. Another limitation of the present study is the use of rigid-wall boundary conditions (\ref{bc}). However, if necessary, the developed formalism can be modified for the cases of open or asymmetric boundary conditions, which is out of scope of the present paper. }

\begin{acknowledgements}
We would like to thank the unknown referee for constructive comments. This work is supported by the BK21 plus program through the National Research Foundation (NRF) funded by the Ministry of Education of Korea, Basic Science Research Program through the NRF funded by the Ministry of Education (NRF-2013R1A1A2012763), NRF of Korea Grant funded by the Korean Government (NRF-2013M1A3A3A02042232), the Korea Meteorological Administration/National Meteorological Satellite Center, (SK, VMN, YJM); and
by the European Research Council under the \textit{SeismoSun} Research Project No.~321141 (VMN).\end{acknowledgements}

\bibliographystyle{apj}
\bibliography{ms}

\end{document}